\documentclass[12pt]{article}
\usepackage{epsfig,pslatex}

\begin{document}

\baselineskip 20pt

\begin{flushright}
{\bf SDU-HEP200705}
\end{flushright}

\vskip 20pt

\begin{center}

{\LARGE{\bf Doubly heavy baryon production at polarized photon collider}}\\
\vspace{20pt}
{\small \it Zhong-Juan Yang\footnote{yangzhongjuan@mail.sdu.edu.cn},
Tao Yao\footnote{yt@sdu.edu.cn}\\
Department of Physics, Shandong University, Jinan, 250100,
P. R. China}

\vspace{20pt}
{\bf Abstract}
\end{center}
\noindent
 We study the inclusive production of doubly heavy baryon $\Xi_{cc}$ at
polarized photon collider. Our results show that proper choice of the 
initial beam polarizations may increase the production 
rate of $\Xi_{cc}$ approximately 10\%.

\vspace{0.3cm}

\vspace{0.3cm}

\vspace{0.5cm}


The doubly heavy baryon production is an important  topic in both experiments 
and theoretical studies. SELEX collaboration has observed the  doubly heavy 
baryon  $\Xi_{cc}$ \cite{Mattson:2002vu,Moinester:2002uw,Ocherashvili:2004hi}.
Many theoretical works have been done in
\cite{Falk:1993gb, Kiselev:1994pu, Berezhnoi:1995fy, Baranov:1995rc, 
Berezhnoi:1998aa, Ma:2003zk, Chang:2006eu, Chang:2006xp, Li:2007vy}. 
However, the production rate and decay width measured at SELEX  are much 
larger than most of the theoretical predictions.
So, it is necessary to study the production 
mechanism more precisely. At the future International Linear Collider (ILC),
backscattered laser light may provide very high-energy photons
\cite{Ginzburg:1982yr}. In 
(un)polarized photon photon fusion, the doubly heavy baryon production is 
possible with certain rates. The process 
$\gamma\gamma \to H_{QQ}X$ may play an important role in the 
production mechanism of doubly heavy baryon.

The production of doubly heavy baryon can be divided into two steps: the 
first step is the perturbative production of a  heavy quark pair. 
For two identical heavy quark system, there are two states 
contributing to the doubly heavy baryon production, one state is with the 
heavy quark pair in $^3S_1$ and color triplet, the other is with the 
pair in  $^1S_0$ and color sextet. 
 The second step is 
the transformation of the heavy quark pair into the baryon, which is 
nonperturbative. Since a  heavy quark has a small velocity 
in the rest frame of the baryon, we can use non-relativistic QCD (NRQCD)
to handle the transformation \cite{Bodwin:1994jh}.
Two hadronic matrix elements are defined for the transformation from the two 
states \cite{Ma:2003zk,Brambilla:2005yk}. In refs.
\cite{Falk:1993gb, Kiselev:1994pu, Berezhnoi:1995fy, Baranov:1995rc, 
Berezhnoi:1998aa} the inclusive 
production of doubly heavy baryons has been studied and only the 
contribution from quark 
pair in $^3S_1$ and color triplet is taken into account. 
According to the discussion in 
\cite{Ma:2003zk},
one should include the contribution from both color triplet and color sextet.
In refs. \cite{Ma:2003zk, Chang:2006eu, Chang:2006xp},  
the production of doubly heavy baryons for $e^+e^-$  and  hadron-hadron   
colliders are studied by considering the contribution from
the two states. The  inclusive production of $H_{QQ}$
at unpolarized $\gamma \gamma$ collider is investigated 
in \cite{Li:2007vy}. In this letter, we investigate  
the inclusive production of
doubly heavy baryon at polarized photon collider. It is found that 
proper choice of the initial beam  polarizations may increase  
the production rate of $\Xi_{cc}$
up to 10\%.

\begin{figure}
\begin{center}
\psfig{file=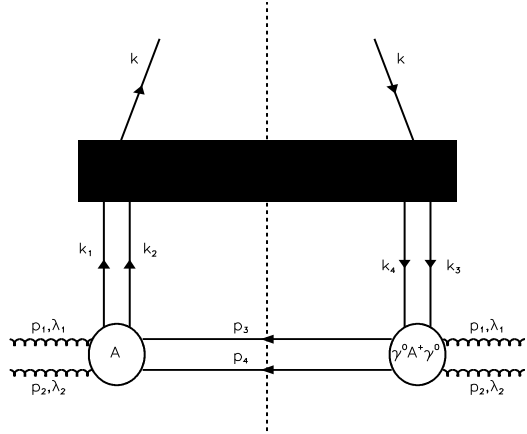, height=8cm, width=10cm}
\caption{Graphic representation for the contribution in eq.(\ref{wen6}), the 
black box represents the Fourier transformed
matrix element, the
dashed line is the cut and $k_4=k_1+k_2-k_3$.}
\label{fey4q}
\end{center}
\end{figure}
\begin{figure}
\begin{center}
\psfig{figure=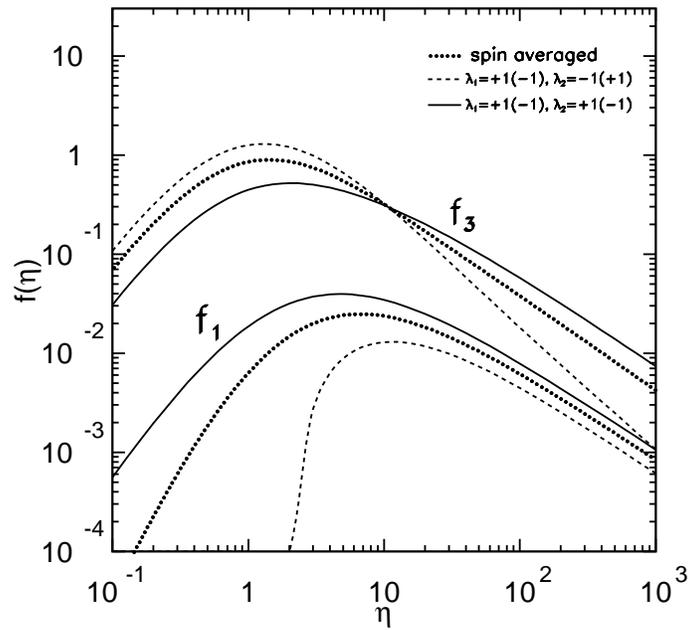,width=10cm}
\caption{The scaling function $f(\eta)$.}
\label{fhnpa}
\end{center}
\end{figure}

The inclusive production of doubly heavy baryon $H_{QQ}$ by photon 
scattering can be described as
\begin{equation}
\label{wen1}
\gamma(p_1,\lambda_1)+\gamma(p_2,\lambda_2) \to H_{QQ}(k)+X ,
\end{equation}
where $p_1$, $p_2$ and $k$ are respectively the momenta of the
corresponding particles, $\lambda_1$ and $\lambda_2$ are the helicities of 
the photons. X is the unobserved state, and we can always divide it into a
nonperturbative part $X_N$ and a perturbative part $X_P$. At leading order,
$X_P$ consists of two heavy anti-quarks $\bar{Q}$.
With  the same notation as that in refs.\cite{Ma:2003zk,Li:2007vy}, one 
can obtain 
the differential cross section for the process of eq.(\ref{wen1}) 
\begin{eqnarray}
\label {wen6}
d\hat{\sigma}(\hat s,\lambda_1,\lambda_2 )&=&\frac{1}{4}\frac{1}
{2\hat{s}}\frac{d^3\bf k}
{(2\pi)^3}\int \frac{d^3p_3}
{(2\pi)^3 2E_3}\frac{d^3p_4}{(2\pi)^3 2E_4}\int \frac{d^4k_1}{(2\pi)^4}
\frac{d^4k_3}{(2\pi)^4}\frac{1}{2}
 A_{ij}(k_1,k_2,p_3,p_4,\lambda_1,\lambda_2)\nonumber\\
&&\cdot \frac{1}{2} (\gamma^0 A^\dagger(k_3,k_4,p_3,p_4,\lambda_1,\lambda_2)
\gamma^0)_{kl}
\int d^4x_1 d^4x_2 d^4x_3 e^{{-ik_1}\cdot{x_1}-{ik_2}
\cdot{x_2}+{ik_3}\cdot {x_3}}\nonumber \\
&&\cdot \langle0|Q_k(0)Q_l(x_3)a^\dagger({\bf k})
a({\bf k})\bar{Q_i}(x_1)\bar{Q_j}(x_2)|0\rangle,
\end{eqnarray}
where $k_1$, $k_2$ denote the momenta of the internal heavy quarks, 
and $p_3$, $p_4$ the momenta of the anti-quarks. 
i, j are Dirac and color indices, and
$Q(x)$ is the Dirac field for the heavy quark. The 
summation over the final state's spins and  colors are implied here, 
and $\hat s=(p_1+p_2)^2$. The factor $1/4$ is because of 
the identical phtons and anti-quarks. $a^\dagger({\bf k})$ is 
the creation operator for $H_{QQ}$ with three momentum {\bf k}.
The contribution of eq.(\ref{wen6}) can be represented graphically by
Fig.\ref{fey4q}.
In the framework of NRQCD, at the zeroth order
of the relative velocity between heavy quarks in the rest frame
of $H_{QQ}$, we can handle  the hadronic matrix element as in \cite{Li:2007vy}.
Then, the cross section for
$\gamma(p_1,\lambda_1)+\gamma(p_2,\lambda_2) \to H_{QQ}(k)+\bar 
Q(p_3)+\bar Q(p_4)+X_N$
process can be expressed as
\begin{equation}
\label{asde}
\hat{\sigma}(\eta,\lambda_1,\lambda_2)=\frac{\alpha^2 \alpha_s^2 e_q^4}
{m_Q^2} [f_1(\eta,\lambda_1,\lambda_2)
\frac {h_1}{m_Q^3}+ f_3(\eta,\lambda_1,\lambda_2) \frac {h_3}{m_Q^3}],
\end{equation}
where $m_Q$ is the mass of the heavy quark,  $\alpha$ and $\alpha_s$ are 
the fine structure constant and strong coupling, $e_q=2/3(-1/3)$ for 
up(down)-type quark,
and $\eta=\hat s/(16m_Q^2)-1$. $ h_1$ ($ h_3$) 
represents the probability
for a $QQ$ pair in  $^1S_0$ ($^3S_1$) state and in the color state of
$6$ ($\bar 3$) to transform into the baryon \cite{Ma:2003zk,Li:2007vy},
and they should be determined by
nonperturbative QCD. Under NRQCD, $h_1$ and $h_3$ are at the same
order. $h_3$ can be related to the non-relativistic wave function at 
the origin, i.e., $h_3=|\Psi_{QQ}(0)|^2$.
The numerical results for the scaling functions 
$f_1(\eta,\lambda_1,\lambda_2)$ and $f_3(\eta,\lambda_1,\lambda_2)$ 
are displayed in fig.\ref {fhnpa}. One can notice 
 $f_1$ and  $f_3$ do not depend on $m_Q$ explicitly\footnote{All the results 
in \cite{Li:2007vy} had left out
a factor 1/4.}.

\begin{figure}
\begin{center}
\psfig{figure=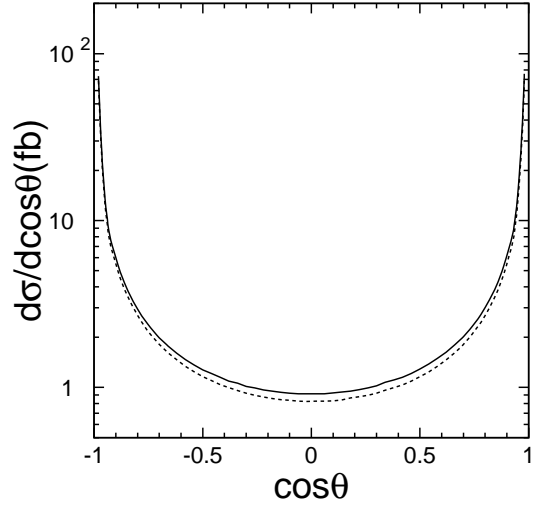,width=8cm}
\caption{$cos\theta$-distributions with $h_3=h_1=|\Psi_{cc}(0)|^2$, 
the solid line for $(P_{e1},P_{e2};P_{L1},P_{L2})=(0.85,0.85;+1,+1)$, 
the dashed for spin averaged.}
\label{efwtg}
\end{center}
\end{figure}

\begin{figure}
\begin{center}
\psfig{figure=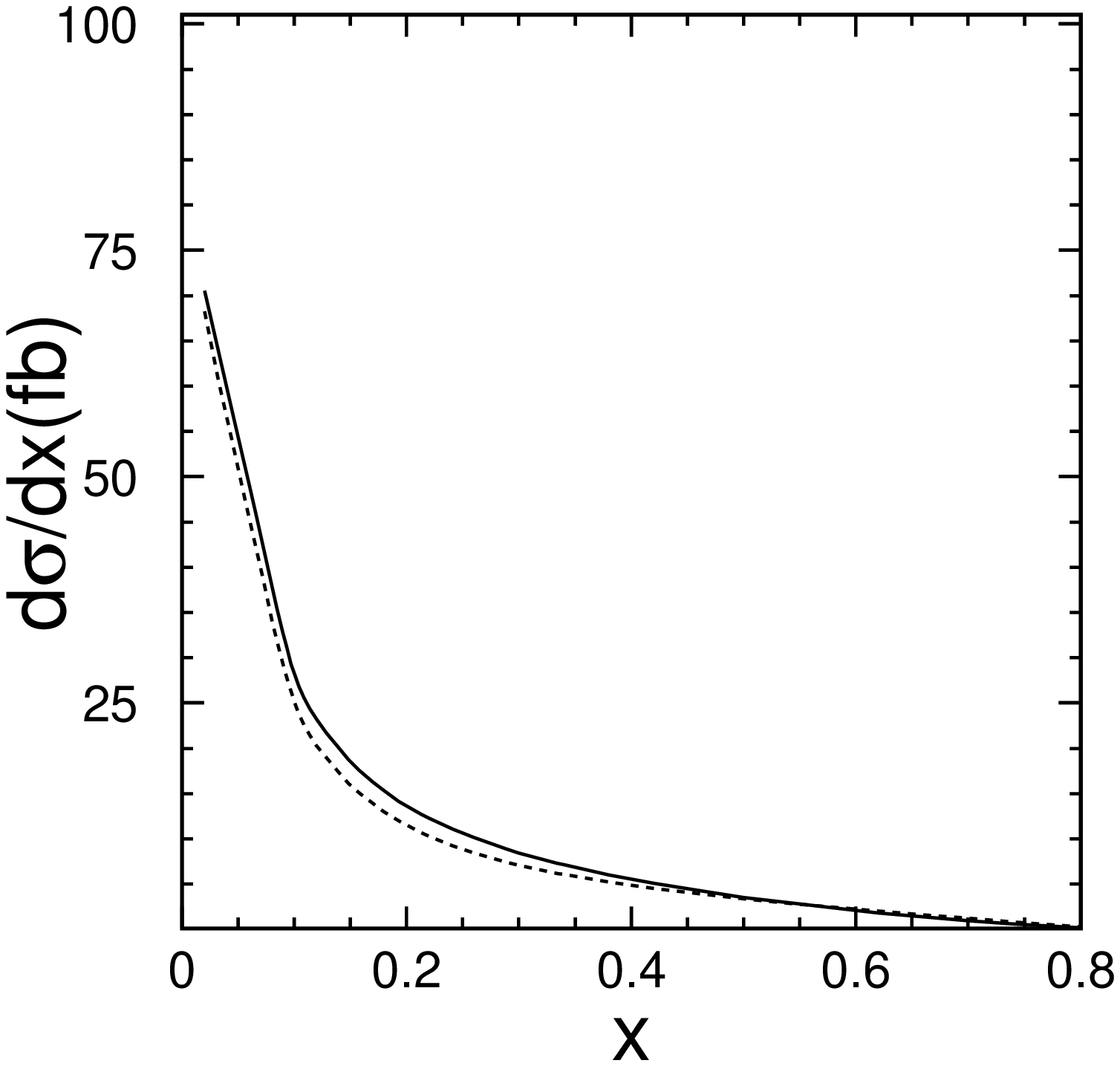,width=8cm}
\caption{Same as Fig.\ref{efwtg}, but for $x$-distributions.}
\label{errdj}
\end{center}
\end{figure}

\begin{figure}
\begin{center}
\psfig{figure=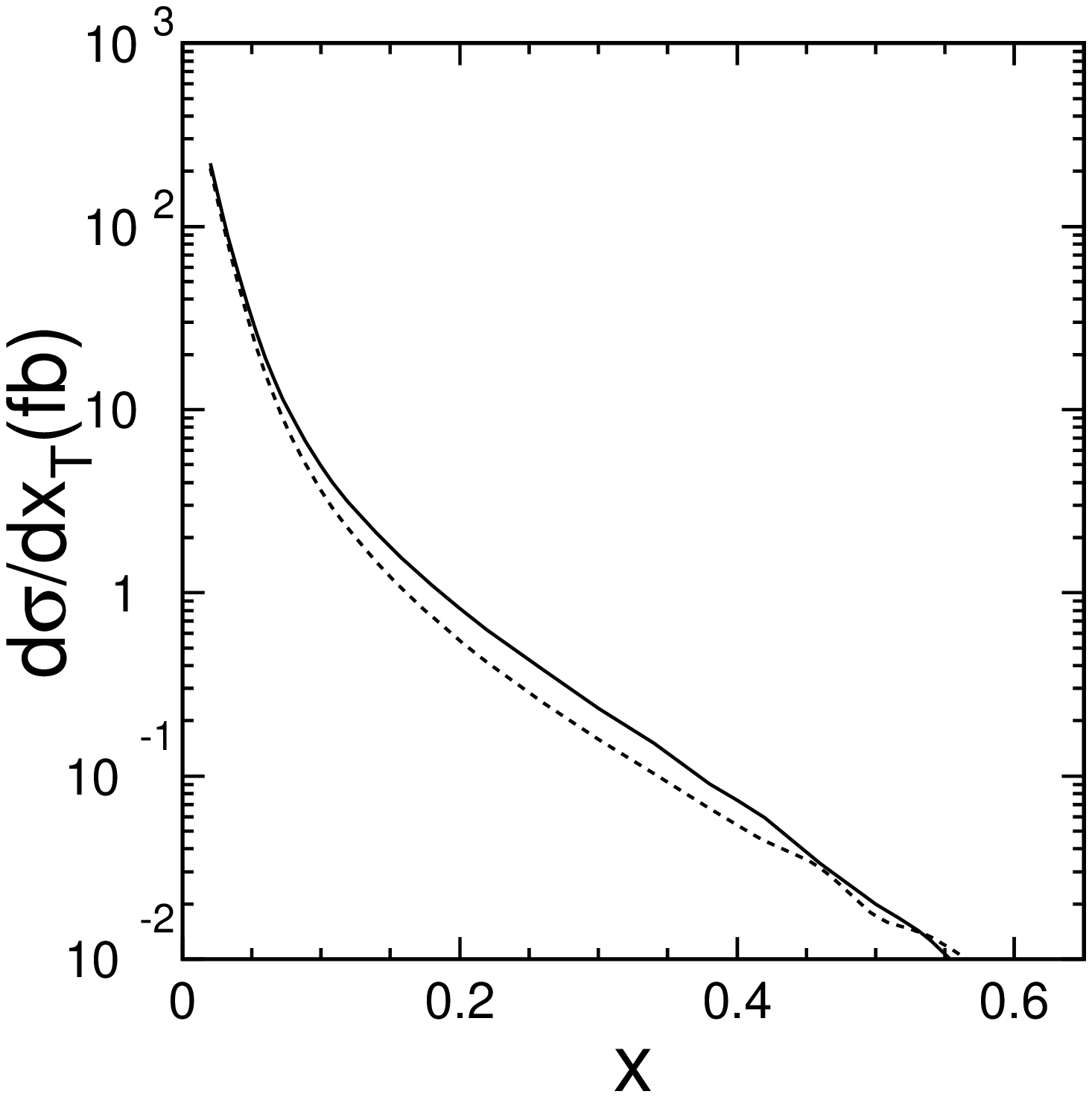,width=8cm}
\caption{Same as Fig.\ref{efwtg}, but for $x_T$-distributions.}
\label{xdew}
\end{center}
\end{figure}

The total effective cross section for doubly heavy baryon production at 
a photon collider can be written as
\begin{equation}
\label{eeq14}
d\sigma (S)=\int_{0}^{y_{max}} dy_1 \int_{0}^{y_{max}} dy_2 f_{\gamma}^e
(y_1,P_e,P_L) f_{\gamma}^e(y_2,P_e,P_L) d\hat{\sigma}
(\hat s,\lambda_1,\lambda_2)
\end{equation}
with the normalized energy spectrum of the photons
\begin{equation}
\label{rre}
f_{\gamma}^e(y,P_e,P_L)={\cal N}^{-1}[\frac{1}{1-y}-y+(2r-1)^2
-P_eP_Lxr(2r-1)(2-y)],
\end{equation}
where $ \sqrt{S}$ is the $e^+e^- $ CMS energy, $\cal N$ the normalization 
factor, $P_e(P_L)$ the polarization of the electron (laser) beam, 
$r=y/(x-xy)$, and $y$ is the fraction of the electron energy
transferred to the photon in the center-of-mass frame. It has the following
range
\begin{equation}
0\leq y \leq \frac{x}{x+1}
\end{equation}
with
\begin{equation}
 x=\frac{4 E_L E_e}{m_e^2},
\end{equation}
where $m_e$ is the electron mass and $E_e$ ($E_L$) the energy of the 
electron (laser) beam. In order to avoid the creation of an $e^+e^-$ 
pair from the backscattered laser beam and the low energy laser beam, 
the maximal value for $x$ is $2(1+\sqrt{2})$. 
In the calculation, we 
adopt the following parameter values
\begin{equation}
E_e=250 GeV, \hspace {0.5cm}E_L= 1.26 eV, \hspace {0.5cm}m_e=5.11
\times10^{-4} GeV.
\end{equation}
The differential cross section $ d\hat{\sigma}(\hat s,\lambda_1,\lambda_2)$ 
is evaluated with polarizations
\begin{equation}
\lambda_i=P_\gamma(y_i,P_e^{(i)},P_L^{(i)}), i=1,2,
\end{equation}
where the function $P_\gamma(y,P_e,P_L)$ is the polarization 
of photons resulting from Compton backscattering with energy fraction $y$, 
which is
\begin{equation}
P_\gamma(y,P_e,P_L)=\frac{1}{f_{\gamma}^e(y,P_e,P_L)\cal N}
\{xrP_e[1+(1-y)(2r-1)^2]-(2r-1)P_L
[\frac{1}{1-y}+1-y]\}.
\end{equation} 
For consistency, we use the same values 
as taken in \cite{Li:2007vy}
\begin{eqnarray}
&&m_c=1.8 GeV,\hspace{2.0cm} \alpha=\frac{1}{137},\nonumber\\
&&\alpha_s(\mu=2m_c)=0.20, \hspace{1.0cm}|\Psi_{cc}(0)|^2=0.039 GeV^3.
\end{eqnarray}
Numerical results for the total effective cross section of $\Xi_{cc}$ 
are given in Table \ref{corr}. One can notice photon polarization is an 
important asset. The choice $(P_{e1},P_{e2};P_{L1},P_{L2})=(0.85,0.85;+1,+1)$
can increase the production rate of $\Xi_{cc}$ by approximately $10\%$. 
One can also find that the contribution from the color 
sextet $Q  Q$ pair is  about $10\%$ of that from the color triplet one 
if $h_1=h_3$. We also calculate the distributions of $cos\theta$, $x$ 
and $x_T$ for $\Xi_{cc}$, which are given in Fig.\ref{efwtg}, \ref {errdj}
and \ref {xdew} respectively. Here, $\theta$ and $x$ are defined in 
$e^+e^-$ CMS, where $\theta$ is the angle between the
moving direction of $\Xi_{cc}$ and that of the beam. $x=2E/
\sqrt{S}$ and $x_T=2P_T/ \sqrt{S}$, with $E$ and $P_T$ the energy
and transverse momentum of $\Xi_{cc}$ respectively.

\begin{table}
\begin{center}
\begin{tabular}{|c|c||c|c|c|}  \hline
\multicolumn{2}{|c||}{\em$h_1(GeV^3)$}& $0$ &$0.039$ &$0.039$ \\ \hline
\multicolumn{2}{|c||}{\em $h_3(GeV^3)$}& $0.039$ &0 &0.039 \\ \hline \hline
&spin averaged&  8.30 (fb)& 0.88(fb)& 9.18  (fb)\\\cline{2-5}
&$(0.85,0.85;-1,-1)$&7.89(fb) & 0.80(fb)& 8.69 (fb)\\ \cline{2-5} 
{\em $(P_{e1},P_{e2};P_{L1},P_{L2})$}&$(0.85,0.85;+1,+1)$&9.04 (fb) 
     & 0.98(fb)& 10.02 (fb)\\ \cline{2-5}
&$(0.85,-0.85;-1,+1)$&7.77 (fb) & 0.83(fb)& 8.60 (fb)\\ \cline{2-5}
&$(0.85,-0.85;+1,-1)$& 8.46 (fb) & 0.91(fb)& 9.37 (fb)\\ \hline
\end{tabular}
\caption{Results for the effective cross section of $\Xi_{cc}$ at
$\sqrt {S}=500 GeV$.}
\label{corr}
\end{center}
\end{table}

To summarize, we investigate the production of the doubly heavy
baryon $\Xi_{cc}$ at polarized photon collider. 
The production rate of $\Xi_{cc}$ can be increased about 10\% with the 
initial beam polarizations  
$(P_{e1},P_{e2};P_{L1},P_{L2})$ $=(0.85,0.85;+1,+1)$ . The enhancement is 
almost 
equal to the contribution from the the color sextet. The precise
measurement of $\Xi_{cc}$ at polarized photon collider will be helpfull
to understand the doubly heavy baryon production mechanism.

\section*{Acknowledgements}
This work is supported in part by NSFC, NCET of MoE and HuoYingDong
Foundation of China. The authors would like to thank Prof. Z.G.Si
(SDU) for his suggestions and helpful discussions, and also thank
all the other members in the theoretical particle physics group in
Shandong University for their helpful discussions.

\end{document}